\documentclass[prl,twocolumn,citeautoscript,superscriptaddress]{revtex4}

\usepackage{times}

\usepackage[dvips]{graphicx}
\usepackage{epsfig}
\usepackage{dcolumn}

\usepackage{amsmath}

\begin{document}
\title{Theory of neutral nitrogen-vacancy center in diamond and its
  qubit application}

\author{Adam Gali} \affiliation{Department of Atomic Physics, Budapest
  University of Technology and Economics, Budafoki \'ut 8., H-1111, 
  Budapest, Hungary}

\begin{abstract}
  The negatively charged nitrogen-vacancy defect (NV)$^-$ in diamond
  has attracted much attention in recent years in qubit and biological
  applications. The negative charge is donated from nearby nitrogen
  donors that could limit or stem the successful application of
  (NV)$^-$. In this Letter, we unambiguously identify the
  \emph{neutral} nitrogen-vacancy defect (NV$^0$) by \emph{ab initio}
  supercell calculations. Our analysis shows that i) the spin state can
  be \emph{selectively} occupied optically, ii) the electron spin state
  can be manipulated by time-varying magnetic field, and iii) the spin
  state may be read out optically. Based on this NV$^0$ is a new hope
  for realizing qubit in diamond \emph{without} the need of nitrogen
  donors.
\end{abstract}

\maketitle


Realization of qubits is of extremely high importance because that can
be used in quantum cryptography, quantum optics and quantum computing.
One of the most promising candidate is the \emph{negatively charged}
nitrogen-vacancy defect (NV$^-$) \cite{duPreez65,Davies76} that can
operate at \emph{room temperature}
\cite{Wrachtrup01,Jelezko02,Jelezko04-1,Jelezko04-2,Epstein05,Hanson06PRL,Childress06,Childress07,Hanson08}.
The negative charge is donated from nitrogen substitutional
(N$_\text{S}$) \cite{Loubser77,Hanson06}.  It has been very recently
shown that the major source of the decoherence of the electron spin of
the negatively charged NV center is the electron spin bath of nitrogen
substitutionals in type-$Ib$ diamond, that can be eliminated only at
low temperature using a giant magnetic field
\cite{takahashi:047601}. This is a serious limiting factor towards the
development for \emph{practical} applications. In addition, it has
been recently demonstrated that nanometer-sized diamond particles
containing NV$^-$ are useful as fluorescent biomarkers for \emph{in
  vitro} imaging applications \cite{Yu05,Chang08}. However, the
requirement of the extra charge on NV defect from N$_\text{S}$ could
be critical in biomarker applications, where the size of the
nanodiamond is reduced to some nanometers in diameter
\cite{Rabeau07}. One solution of this problem would be to apply the
paramagnetic \emph{neutral} NV defect (NV$^0$) which produces strong
photoluminescence at 2.156~eV (similar to NV$^-$) and it \emph{does
  not require an extra electron}. The NV$^-$ center is well identified
\cite{Loubser77} while its strong connection to its neutral
counterpart is well established experimentally
\cite{Davies92,Mita96,Kennedy03}. However, the overall knowledge about
NV$^0$ is scarce.  Very recently, an $S=\frac{3}{2}$ electron
paramagnetic resonance (EPR) center has been found in photo-excited
diamond doped by $^{15}$N \cite{Felton08}. A sizeable $^{15}$N
hyperfine constants have been detected in the EPR measurements and it
was proposed that the signal was originated from one of the excited
states of NV$^0$ \cite{Felton08}. We emphasize that the identification
of the EPR signal of NV$^0$ is a \emph{key step} in order to apply it
for qubit, and to trace it magnetically in biomarker applications.

In this Letter we i) unambiguously \emph{identify} the EPR signal of
NV$^0$ ii) provide a detailed spin density distribution around the
defect; the results indicate that the defect states are well-localized
and the electron spin can be decoupled from the spin bath of $^{13}$C
nuclei iii) analyze the electronic structure by group theory
explaining the photo-excitation of NV$^0$; this shows that the
$M_\text{S}=\pm 1/2$ sublevels are \emph{selectively} occupied in the
photo-ionization process and appropriate microwave magnetic field can
be applied to choose also selectively either $M_\text{S}=+3/2$ or
$M_\text{S}=-3/2$ which finally can be used as qubit iv) we propose
that the $M_\text{S}=+3/2$ or $M_\text{S}=-3/2$ states may be read out
optically during the emission process from the excited state to the
ground state. This gives a hope for realization of qubit by NV$^0$
defect.


We employed density functional theory calculations with local density
approximation (LDA) using a large, 512-atom simple cubic diamond
supercell. We used two different codes: the geometry of the defect was
optimized with the {\sc VASP} code~\cite{Kresse96} while the hyperfine
tensors of NV$^0$ were calculated by the {\sc CPPAW}
code~\cite{CPPAW}. Both {\sc VASP} and {\sc CPPAW} codes apply the
all-electron PAW method and plane wave basis set. We used a cut-off of
30~Ry and $\Gamma$-point for k-point sampling. Other details and
references can be found in our previous publication \cite{Gali08}
where we could successfully describe the negatively charged NV center
and can reproduce well the experimental hyperfine data.


The NV defect has C$_{3v}$ symmetry, if no reconstruction occurs. We
use defect-molecule and group theory analysis below as for NV$^-$
(c.f. \cite{Gali08} and references therein).  Four dangling bonds
point to the vacant site resulting in $a_1(1)^2a_1(2)^2e^1$
configuration for the neutral defect, where $a_1(2)^2$ and the double
degenerate $e^1$ appear in the fundamental band gap. In this
configuration, the system is basically Jahn-Teller unstable because
the degenerate state is only partially filled by electrons. The static
Jahn-Teller effect would result in C$_{1h}$ symmetry of the defect. PL
studies indicate that the system has C$_{3v}$ symmetry ($^2E$ state)
with possibly exhibiting dynamic Jahn-Teller effect that inhibits the
EPR detection of the $^2E$ ground state \cite{Davies79}. The C$_{1h}$
symmetry configuration and the $^2E$ ground state of C$_{3v}$ symmetry
can be described by single Slater-determinants, so we could address
this issue directly at LDA level. The lowest excitation can be
obtained by promoting one electron from $a_1(2)$ level to $e$ level
resulting in $a_1(1)^2a_1(2)^1e^2$ configuration. The possible excited
states of $a_1(1)^2a_1(2)^1e^2$ configuration have $^2A_1$, $^4A_2$
and $^2E$ multiplets. The $M_\text{S}=\frac{3}{2}$ state of $^4A_2$
multiplet can be described by a single Slater-determinant by simply
aligning all the electrons spin-up on the $a_1(2)$ and $e$ levels,
thus the spin density can be determined by usual LDA calculation. We
note that no Jahn-Teller effect occurs for $^4A_2$ state, so we
considered only C$_{3v}$ symmetry for this state in the
calculations. We focused our research on the ground state of C$_{1h}$
and C$_{3v}$ symmetries, and the $^4A_2$ excited state that are
relevant in the recent EPR study \cite{Felton08}. We note here that
the nitrogen dangling bond is hybridized into $a_1(2)$ but not into
the $e$ defect states, therefore, negligible spin density is expected
for the $^2E$ ground state but considerable spin polarization may be
expected for the $^4A_2$ excited state.


First, we investigated the ground state of NV$^0$. Since Jahn-Teller
effect can occur, we distorted the symmetry to C$_{1h}$, and allowed
the atoms to relax to find the energy minimum within LDA. We found
that the system conserves its C$_{1h}$ symmetry and it does not form
the C$_{3v}$ configuration. The single occupied $e$-level in the gap
is split by 0.3~eV resulting in an occupied $a'$ state and an
unoccupied $a"$ state of the spin-up electrons.  The neighbor N and C
atoms relaxed outward from the vacant site. The nitrogen atom remained
very close to the $\langle 111\rangle$ axis even in C$_{1h}$ symmetry,
while one of the carbon atoms is considerably closer to the vacant
site than the other two. Next, we constrained the system to preserve
the C$_{3v}$ symmetry during the geometry optimization in order to
calculate the total energy of $^2E$ state. We found that it is about
0.09~eV higher in energy than the Jahn-Teller distorted one. Then, we
allowed the system to relax without symmetry constraints starting from
the optimum C$_{3v}$ geometry. The atoms automatically relaxed to the
C$_{1h}$ symmetry. This result indicates that C$_{3v}$ symmetry is a
local maximum. This finding seems to contradict the PL spectrum which
shows $^2E$ ground state within C$_{3v}$ symmetry. However, we carried
out quasi-static calculations at 0~K, and we did not take the vibronic
states and temperature effects into account. This result shows clearly
a multi-valley potential surface for the ground state of this defect
as was already hinted by Davies \cite{Davies79}. Our calculated
Jahn-Teller energy (0.09~eV) is very close to the estimation of Davies
(0.14$\pm$0.07~eV) based on the luminescence measurements
\cite{Davies79}. There are three equivalent C$_{1h}$ configurations
around the vacant site rotated by 120 degrees about the C$_{3v}$-axis
with each possessing the global minium. At LDA-level, we can estimate
the upper limit of the energy barrier between the three global minima,
that is $\approx$0.09~eV.  There may exist more favorable path between
these global minima than through the C$_{3v}$ configuration, so the
actual barrier energy may be even lower than that. Due to the strong
C-C bonds in diamond $\leq$0.09~eV energy difference can be gained
even at very low temperature because this is about the zero point
energy of the phonons with the highest energy ($\leq$0.083~eV). This
results in a motional average of the single electron between the three
equivalent C$_{1h}$ configurations, showing an effective C$_{3v}$
symmetry. It was speculated \cite{Davies79,Felton08} that the dynamic
Jahn-Teller effect is responsible for the missing EPR signal of the
$^2E$ ground state. Our calculations support this assumption.

We also investigated the $^4A_2$ excited state. This state has indeed
much higher energy by about 0.86~eV compared to that of the
low-symmetry ground state. 
We note that the LDA total energy differences
should not be directly compared to the measured transition energies as
previously discussed in Ref.\ \onlinecite{Gali08}. 
In this state the $a_1(2)$ defect level is polarized, thus the
nitrogen atom is polarized in contrast to $^2E$ state. 
%
This results in considerable rearrangement
of the atoms around the vacancy, namely, the nitrogen moves closer to
the vacant site. We calculated the hyperfine tensors of atoms for the
optimum geometry and compared to a recently found EPR center as
explained in the introduction \cite{Felton08} (see
Table~\ref{tab:hyperf}).
\begin{table}
\caption{\label{tab:hyperf} The calculated principal values of the
  hyperfine tensor (columns 2 to 4) compared to the known experimental
  data (columns 5 to 7) in MHz. The experimental data on $^{15}$N is taken from
  Ref.~\onlinecite{Felton08}. The $^{13}$C hyperfine signal could not
  be resolved in that experiment.}
\begin{ruledtabular}
\begin{tabular}{ccccccc}
Atom & A$_{11}$ & A$_{22}$ & A$_{33}$ & A$^\text{exp}_{11}$ &
  A$^\text{exp}_{22}$ & A$^\text{exp}_{33}$\\ \hline
$^{15}$N           & -23.4     &    -23.4    &    -39.0
                   & -23.8(3)  &   -23.8(3)  &   -35.7(3)  \\  
$^{13}$C(3$\times$)   &  60.6&    61.0&   126.2 &   &  &  \\
\end{tabular}
\end{ruledtabular}
\end{table}
The agreement between the calculated and measured hyperfine signal is
excellent. In addition, the calculated binding energy of the NV$^0$
complex ($\approx$3.0~eV) shows a high thermal stability, and its
(-/0) occupation level is at about 2.0~eV above valence band edge, in
line with the experimental observations \cite{Mita96}. Thus,
we identify the EPR signal of NV$^0$. We provide the hyperfine data of
the three nearest carbon atom near the vacant site that has the
largest hyperfine interaction. These carbon atoms may be measured by
future EPR experiments when the signal to noise ratio can be reduced
there.

After identification of NV$^0$ defect we discuss its possible role in
spin physics. NV$^-$ was successfully used to realize qubits
\cite{Childress06,Childress07}. Second order correlation and EPR
measurements were employed to detect individual NV$^-$. The resulting
spin-echo signals show a rapidly oscillating function enveloped by a
more slowly oscillating function~\cite{Childress06}.  The authors of
Ref.\ \onlinecite{Childress06} proposed a theory to explain this
signal, and they concluded that the fast modulation frequency is due
to the effective magnetization density of the electron spin felt by
the $^{13}$C nucleus, which is the same as the hyperfine interaction.
If too many $^{13}$C nuclei are involved in the process that will lead
to fast decoherence of the spin-echo signal \cite{Childress06},
therefore, the knowledge of the spin density distribution is crucial.
The situation is complex for the ground state as was discussed above.
At the measurement temperature an effective C$_{3v}$ symmetry may be
detected as the motional average of the three equivalent C$_{1h}$
configurations. It is difficult to handle this situation by
quasi-static simulation. We simulate this average simply by taking the
optimum $^2E$ ground state within C$_{3v}$ symmetry.
We found that the overall picture is very similar to what was found
for NV$^-$ \cite{Gali08}. In the ground state the spin density is
mainly localized on the three nearest neighbor carbon atom of the
vacant site, and the spin density decays fast as a function of the
distance from the vacant site. From this point of view, a potential
qubit of NV$^0$ would be weakly coupled to its environment and,
therefore, it might present the right coherence properties like
NV$^-$.

However, the localized spin density is not the only requirement to
produce qubits. One needs to generate a superposition state and
read-out the qubit states. In the case of NV$^-$ the $M_\text{S}=0$ of
the triplet \emph{ground state} can be optically pumped which has much
smaller hyperfine interaction with the proximal $^{13}$C nucleus than
the $M_\text{S}=1$ state \cite{Childress06}. This effect was
responsible for the collapse and revival of the electron spin coherent
state. The $M_\text{S}$ states could be read-out also optically after
the measurements by using the fact that they have different
fluorescence rates \cite{Childress06}.

\begin{figure}
\centering    
\includegraphics[keepaspectratio,width=8.0cm]{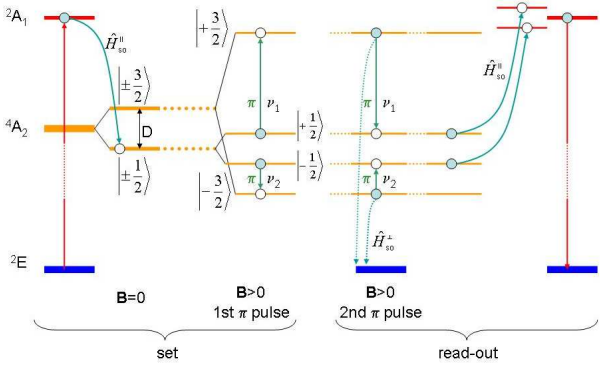}
\caption{\label{fig:levels}(Color online) The process of the
  manipulation of a qubit in NV$^0$ defect.  Straight red arrow:
  radiative recombination. Curved green arrow: spin-orbit coupling
  ($\hat H_\text{SO}$) possibly mediated by phonons. Grey dotted
  arrows represent very weak interaction. Blue straight line:
  microwave alternating magnetic field pulse. Filled(empty) circle:
  initial(final) state of the electron. Setting the qubit state: i)
  excitation from the $^2E$ ground state to the $^2A_1$ excited state
  ii) spin-flip to the appropriate $^4A_2$ states iii) setting either
  $M_\text{S}=3/2$ or $M_\text{S}=-3/2$ state by $\pi$ pulse. Readout
  the qubit state: i) $\pi$ pulse to go back to $M_\text{S}=\pm1/2$
  states ii) spin-flip to the $^1A_1$ state iii) radiative
  recombination to the $^2E$ ground state. We show the fine structure
  of $^4A_2$ states in the absence of external magnetic field
  ($\mathbf{B}$=0) (2nd column) and at $\mathbf{B}>0$ with typical EPR
  conditions in the next columns. $D\approx 1685$~MHz fine structure
  constant was determined by EPR \cite{Felton08}. The figure does not
  show the true scale for the sake of clarity. }
\end{figure}

In NV$^0$ defect the optical pump can change the $S=\frac{1}{2}$ state
to $S=\frac{3}{2}$ \emph{excited state}, and by switching off the
light excitation this can be transformed back to the $S=\frac{1}{2}$
state. The nature of the spin-flip process will be discussed
shortly. The group theory analysis tells us that no
spin-orbit coupling arises for the $^4A_2$ excited state. However, the
spin-spin interaction is active, which can be given by the following
effective spin-Hamiltonian for this particular system:
$\hat{H}_\text{SS} = D' (S_z^2 - 5/4)$, where $D'$ is the zero-field
constant and $S_z$ has the $M_\text{S}$ eigenvalue.
So, the $4\times$ degenerate $^4A_2$ state will split to two double
degenerate states due to spin-spin interaction resulting in the lower
lying $M_\text{S}=1/2;-1/2$ and the upper lying $M_\text{S}=3/2;-3/2$
levels (see Fig.~\ref{fig:levels}). These levels are separated by
$2\times D' = D$. $D\approx 1685$~MHz was experimentally measured by
EPR \cite{Felton08}. Acording to Davies PL analysis \cite{Davies79}
$\sim 2.2$~eV excitation occurs between $^2E$ ground state to the
$^2A_1$ excited state. We propose that the $M_\text{S}=1/2;-1/2$
electron states of the $^2A_1$ excited state can relax to the
$M_\text{S}=1/2;-1/2$ sublevels of $^4A_2$ state with a finite
probability instead of relaxing back to the ground state. While the
spin-orbit interaction is not active for $^2A_1$ state itself the
\emph{axial spin-orbit interaction couples the $^2A_1$ states
  selectively with the $M_\text{S}=1/2;-1/2$ $^4A_2$ states}. That is
the source of the spin-flip process which may be further mediated by
phonons in order to satisfy the energy conservation of this
transition. Indeed, the treshold excitation energy of the EPR signal,
2.2(1)~eV \cite{Felton08}, is larger than the ZPL energy of 2.156~eV,
and it has the highest intensity using excitation energy of about
2.5~eV. The probability of the relaxation process between the original
$^2E$ and $^1A_1$ ($S=\frac{1}{2}$) states should be higher than for
the original $^4A_2$ ($S=\frac{3}{2}$) state. One may estimate from
the known data of the NV$^-$ center \cite{Manson06} that the duration
of the direct transition process is about 10~ns, while for the spin
flip process it could be about 30~ns. Therefore, the duration of the
optical pumping can be relatively long in order to arrive at
$M_\text{S}=1/2;-1/2$ sublevels of $^4A_2$ state from the $^2E$ ground
state. The typical EPR condition for $^4A_2$ state is shown in
Fig.~\ref{fig:levels}. The small external constant magnetic field
($\mathbf{B}$) will split the $M_\text{S}=1/2;-1/2$ sublevels lowering
the energy of $M_\text{S}=-1/2$ state.  Finally, we predict that
during the optical pumping the $M_\text{S}=1/2;-1/2$ sublevels of
$^4A_2$ state will be \emph{selectively} populated with somewhat
higher probability for $M_\text{S}=-1/2$ because it is lower in
energy. Then, the varying microwave magnetic field induce the
transitions $M_\text{S}=\pm1/2 \leftrightarrow \pm3/2$ in the EPR
measurements. This scenario can explain all the photo-EPR findings in
Ref.\ \onlinecite{Felton08}.

The question arises what kind of entity can be used as a qubit from
NV$^0$ defect. The first choice it to use simply the $S$-state. By
optical pumping the $S$-state of NV$^0$ defect can be transformed from
1/2 to 3/2, and by switching off the light it can be transformed back
from 3/2 to 1/2 with $t\ll 1$~s \cite{Felton08}. However, it is not
probable that coherent state can be achieved for the ground state as
we showed that it exhibits a dynamic Jahn-Teller effect causing a
\emph{rapidly varying effective magnetic field} for the $^{13}$C
nuclei around the NV$^0$ defect. Furthermore, the time-averaged
effective hyperfine interaction of both states are very similar. This
is also disadvantageous.

Another possibility is to use the $M_\text{S}$ sublevels of the
$^4A_2$ state as qubit. As explained above by optical pumping one can
select the $M_\text{S}=1/2;-1/2$ sublevels of $^4A_2$ state of the
NV$^0$ defect with \emph{almost equal} probability.  That is not
applicable for qubits.  However, one can selectively set either
$M_\text{S}=+3/2$ (with energy of $h\nu _1$) or $M_\text{S}=-3/2$
($h\nu _2$) states by applying a $\pi$ pulse \cite{Childress06} to
induce the EPR transition ($\Delta M_\text{S}= \pm 1$) (see
Fig.~\ref{fig:levels}). \emph{The $M_\text{S}=\pm 3/2$ states may be
  used as qubit.} It is clear in the above mentioned scenario that
$\nu _1$ can \emph{only} be associated with $M_\text{S}=+3/2$ while
$\nu _2$ \emph{only} with $M_\text{S}=-3/2$. We would like to
emphasize that these metastable $M_\text{S}=\pm3/2$ states are
extremely long living ($>$1~$\mu$s) because i) the $^4A_2$ state is
the only $S=3/2$ state, so there is no way for radiative recombination
for this state ii) the $M_\text{S}=\pm3/2$ states are \emph{not
  coupled to the $^2A_1$ excited state at all that hinders the
  recombination of these states to the $^2E$ ground state via $^2A_1$
  excited state.} This is unique compared to the NV$^-$ center
\cite{Manson06} and suggests a longer lifetime of these metastable
states than for the measured lifetime of the singlet metastable state
in NV$^-$ center (300~ns) \cite{Manson06}. We note that the
$M_\text{S}=\pm3/2$ states are very weakly coupled to the $^2E$ ground
state by \emph{non-axial} spin-orbit interaction. However, the
\emph{non-axial} spin-orbit interaction is very small (as assumed for
NV$^-$ center \cite{Manson06}) and \emph{it is too far in energy in
  order to mediate this process by phonons}. So, the probability of
this decay is very small ensuring the very long lifetime of these
states. The coherent state between these $M_\text{S}=\pm3/2$ states
and the proximate $^{13}$C nuclei can certainly be maintained similar
to NV$^-$ \cite{Childress06}. The readout process could be very
simple: by applying a $\pi$ pulse again the $M_\text{S}=\pm3/2$ states
scatter to $^4A_2$ $M_\text{S}=\pm1/2$ states that through the
spin-orbit coupling can go back to the $^2A_1$ excited state, finally
by radiative recombination to the $^2E$ ground state. Therefore, the
spin qubit state can be readout optically. This process can be tuned
by applying the appropriate constant magnetic field to split the
levels and the microwave magnetic field ($\pi$ pulse) to induce
transitions between the levels.

In this work we investigated the \emph{neutral} nitrogen-vacancy
defect in diamond in detail by \emph{ab initio} LDA supercell
calculations. We showed that the defect indeed shows the dynamic
Jahn-Teller effect for the ground state. We identified the recently
found EPR center \cite{Felton08}, as the $^4A_2$ excited state of the
\emph{neutral} nitrogen-vacancy defect. That EPR center can be used
to trace the NV$^0$-contained nanodiamonds magnetically. We found that
NV$^0$ is a promising candidate for realizing qubits in diamond
\emph{without the need of nitrogen donors}.

AG acknowledges support from Hungarian OTKA No.\ K-67886. The fruitful
discussion with Jeronimo Maze is appreciated.


\end{document}